# Near-Field Optical control of Doughnut-Shaped Nanostructures

[1,2]A. M. Dubrovkin, [1]R. Barillé, [3]E. Ortyl, [3]S. Zielinska,

[1]LUNAM Université, Université d'Angers/UMR CNRS 6200, MOLTECH-Anjou,
2, bd Lavoisier, 49045 Angers, (France).
[2] Centre for Disruptive Photonic Technologies, Nanyang Technological University,
50, Nanyang Avenue, Singapore 637371, (Singapore).
[3]Department of Polymer Engineering and Technology, Wroclaw University of Technology,
Faculty of Chemistry, 50-370 Wroclaw, (Poland)

E-mail: regis.barille@univ-angers.fr, alexandre.dubrovkin@univ-angers.fr

The application of a local near-field optical excitation can be used to control step-by-step the reshape of individual doughnut-shaped azopolymer nano-objects by varying the time of illumination demonstrating its promising performance as a functional nano-object. The possibility to provide both photoinduced reshaping opens a way to the fundamental study of size-dependent scaling laws of optical properties, photoinduced reshaping efficiency and nanoreactor or nanoresonator behavior at nanometer scale. As an example the nano-object is used to self-assembly polystyrene nanospheres in a supraball.

1. Introduction

Light-induced molecular movements in azopolymers have attracted much attention since these last 20 years and are recognized as a possible technology for a broad range of fundamental and applied research [1 - 2]. Many applications take advantage of the photo-induced isomerization where molecules switch from the more stable *trans* form to the less stable *cis* form. One of the most interesting phenomena associated with the photo-isomerization process is massive photoinduced macroscopic motions of the polymer chains leading to physical surface deformation of the material well below the glass transition temperature. Nanostructures induced by the photomechanical properties of azopolymers can be optimized and reshaped by tailored light fields to obtain perfect performances [3 - 5].

Current research in topics of near-field optics, such as optical nanoantennas and localized surface plasmon resonances, has considerably benefited from the unique property of photoinduced azopolymer nanostructures exhibiting polarization and localization of photoinducing near-fields [6]. A recent study has shown the possibility to create induced spiral surface reliefs. The azopolymer thin film was exposed to a focused Laguerre-Gauss like vortex beam and the pattern results from a transfer of the wavefront structure of the writing beam. To date, only few examples show a broad range of one, two and three dimensional azopolymer nanostructures grown by chemical synthesis with near-field illumination and, particularly, with a scanning near-field optical microscope (SNOM) lithography [7]. The macroscopic movement of azopolymer chains by near-field probes was demonstrated [8] and the inscription of local surface relief gratings as well [9].

In this work we experimentally demonstrate a possibility of further flexible growth and reshaping of doughnut-shaped nanostructures by laser illumination through a SNOM probe, what represents their efficient uses as functional nano-objects. Doughtnut-shaped nano-structures (including nano-doughnuts, nano-rings, nano-toroids and particular cases of nano-wells), are rich objects for fundamental study of light localization at nanoscale and have also grabbed attention recently due to their efficient uses in nanoplasmonics, photochemistry, nanorectors and sensors.

Benefited from the high spatial resolution, the near-field technique lets us reshape nano-doughnuts separately. Furthermore, we show an experimental model demonstrating a possible application of doughnut-shaped azopolymer nanostructures as nano-attractors for aggregation of nanospheres.

2. Experimental results

Polymer thin films are made from a highly photoactive azobenzene derivative containing heterocyclic sulfonamide moieties (MB2I) [10]. The thin films are prepared by dissolving 75 mg of azo-polymer in 1 ml of THF (Tetrahydrofuran) and spin-coated on clean glass substrate. The film thickness was determined with a Dektak profilometer and was around 550 nm. The molecular mass of the polymer determined by GPC was between 14 000 and 19 000 and the glass transition temperature (Tg) was 71° C. The method chosen to create the initial patterns on the surface of the azopolymer thin film is solvent induced dewetting. Randomly placed nano-holes with a mono dispersed diameter with different depths are created. The illumination of the surface with a Xenon lamp during 30 minutes allows the creation of doughnuts with a typical volume of 0.065 $\mu m^3$. The figure 1 shows the details of the surface with nano-doughnuts produced on the surface after white light illumination. Several measurements or different nano-doughnuts have shown that typical heights of a doughnut rim above the film surface vary from 50 nm to 150 nm. All presented informations lead to the natural classification of observed doughnut-shaped structures as nano-doughnuts and nano-wells. Nano-doughnuts represent structures with a central hole lying near the film surface while nano-wells have greater depth.

In order to show a height distribution of photoinduced nano-doughnuts typical film topographies (far from the edge) was studied using scanning electron microscopy (SEM) and SNOM (shear-force mode). A commercial SNOM (Smena, NT-MDT) utilizing a laboratory-made bare tapered optical fiber as a probe was used to achieve a control of the geometry of the photoinduced nanostructures.

Photoinduced shape transformation ability determines a significant advantage of using azopolymer nanostructures as functional materials. Besides the wide range of possible geometrical sizes during the fabrication, azopolymer doughnut-shaped nano-objects must exhibit a property of flexible reshaping under a light illumination. To show this property in action, we use the near-field microscopy technique to deliver the laser radiation into a single nano-well. The SNOM set-up mentioned above performs an experiment in the configuration where the light is sent through the probe to the sample surface in the near-field. Experimentally, at first we obtained a SEM image of the previously prepared photoinduced nano-well (Figure 1). Nano-wells are almost separated from surrounding nanostructures by an average distance greater than their diameters giving an opportunity to use them as a single object. The average distance between the doughnuts is 1.3 µm. The figure 2 shows a three dimensional view of an individual photoinduced doughnut shaped structure obtained on the surface. We note the well preserved symmetry of the tore with a structure of high wall with a high of about 100 nm, a diameter of 0.6 μm and a dimension of the tore of about 0.4 μm. Then we carried out several sequenced iterations consisting in: i) placing of the fiber in the center of the nano-well, ii) illuminating it with the laser radiation during a certain time, shutting the laser radiation and iii) scanning the surface topography again. The Figure 3a-c shows results of three described iterations after 1, 2, and 5 minutes of illumination with the probe. The total illumination time during the whole experiment is done by steps of 10 min. For a better representation we also plotted contour images of the nano-well upper part with rescaled color bars. It is clearly seen that the original symmetry of the nano-well has been changed with the illumination time. The initial state has a small asymmetry and has a small difference from a circular case (1.18 ellipticity factor). This slightly elongation could be attributed both to the real shape of the selected nano-well and to a standard probe-size artifact in topography imaging. The nano-well shape is preserved during the first 5 min of illumination. After 5 additional minutes of local illumination with the laser, the nano-well is changed into two separate semi-rings around the central hole similar to a nano-resonator made with two semi-circular mirrors (Figure 3d). This behavior can be attributed to the complex polarization state of the light in the near-field of SNOM probe. The molecules are aligned perpendicularly to the incident polarization. This effect will distort the photo manipulation of the nano-doughnut. The variation of the two axial components (x or y) as a function of time is not similar. The radius R of the doughnut-shaped structure is defined in the figure 4. After illumination the doughnut's radius shows a nonlinear evolution (fig. 5a). A second order polynomial can fit the variation of R as a function of time. We note that measurements of the evolution of widths and heights of the nano-structure are almost linear as a function of time but with a different rise between x and y evolutions leading to an ellipticity. Calculated ellipticity factors after the 1st, 3rd, 5th and 10th minutes of the illumination are 1.30, 1.38, 1.54 and 1.73 respectively (figure 5b). A larger scale topographic image of the film around the illuminated nano-doughnut after the experiment does not exhibit shape changes of the surrounding nano-objects meaning that the individual transformation of one nano-object does not lead to modifications of the surrounding nano-objects. Due to a large ratio of surface area to volume in the doughnut-shaped structure the behavior of the surface becomes a big factor in characterizing and controlling the mechanical behavior of the 1D element at the nanoscale. We study the modification of the circular nano-ring under uniform action with the light as a remote stimulus. We make use of the classical solution for buckling a circular ring under pressure coupled with the surface elasticity theory to determine the values of the mechanical properties when the characteristic dimensions of the nanostructure are in the nanoscale. We consider a nano-ring and we assume that both the bulk and the surface of the tore are homogeneous, isotropic and linear elastic (figure 4). We denote the radius of the circle formed by the center line as *R*. Let *θ* be the angle between the radius drawn from the center of the circle to a point on it and a chosen radius. We take the positive displacement *w* to be directed radially outwards. The displacement *v* is directed along the tangent of the circle in the sense in which *θ* increases. Owing to the displacements w and v (fig. 4), the strain ε of the center line of the ring at any point is given by:

$$\varepsilon = \frac{w}{R} + \frac{1}{R}\frac{dv}{d\theta}$$

and the curvature change of the ring element:

$$\kappa = \frac{1}{R^2}\left(\frac{d^2w}{d\theta^2} - \frac{dv}{d\theta}\right)$$

After illumination of the nanostructure with the SNOM during a timescale of 10 minutes we calculate a strain measurement of ε = 35 % and a change of curvature of - 0.16 rad/m. For comparison the wrinkling of ultrathin polymer films is obtained with a strain of 20 % [11]. These two values show the great possibility of light to dynamically photoinduce a mechanical change of the azopolymer structures.

The study demonstrates a possibility of locally photoinduced modifications of separately fabricated doughnut-shaped nanostructures with a large value of mechanical changes. The local modification can be reversibly changed to come back to the initial state by heating the sample above the glass transition temperature.

Smart nano-wells with controlled volume can be exploited for chemical reactor for photo-catalytic and enzyme reactions [12]. Further chemical contrast studies could be implemented with a nano-pipette where the surface chemistry surrounding the nanowell would be different from the surface chemistry within the nanowell. The nanowell size will be adjusted and spatially used for recognition of chemical substance.

Finally, we demonstrate a possibility of self-organization of polystyrene (PS) nanospheres on the patterned film. An aqueous dispersion of monodispersed colloids containing 250 nm nanospheres was injected on the typical area of the film containing doughnut-shaped nanostructures with 450 nm diameters. A flow of the aqueous solution on the surface was used. The sample was inclined to let the liquid moving slowly. The flow rate had no significant influence on the yield of the self-assembly clusters. The sample is then slowly shacked during the evaporation process. When the liquid dewetted from the bottom surface, the capillary force was sufficiently strong to push the colloidal particles into the template holes, and left almost no colloids on the top surface of the template. If the concentration of colloidal dispersion was high enough, each template would be filled with the maximum number of colloidal particles as determined by geometrical confinement. The colloidal beads within each hole tended to be in physical contact as a result of attractive capillary force caused by solvent evaporation.

After water evaporation we expected to observe doughnuts with nanospheres inside. Figure 6 shows a SEM image of nanospheres trapped into the doughnut-shaped nanostructures. Nanospheres appeared to be placed in doughnuts. Despite the fact that self-organization of nanospheres on previously tailored substrates is a well-known phenomenon [13], it needs efforts if we want a structural control provided by the vertical or the horizontal dimension of the template holes. A demonstration of nanosphere arrangement represents a model experiment which shows a possibility to use created doughnut-shaped nanostructures as a nanoenvironment (nanoreactor) for particles trapped inside. The nano-doughnut can also serve as a template to grow any nanoparticles along the ring.

Final packing of the PS nanospheres leads to the formation of higher-order clusters, or supraparticles, namely, supraballs, as shown schematically in Figure 4. When the number of nanospheres inside the droplets is relatively small, the resultant self-assembled structures will be higher-order clusters or small supraparticles with intermediate packing ordering. These structures have been explained in greater detail [14]. We roughly estimated the number of constituent PS particles for the supraparticles as twelve of particles, from observation of the scanning electron micrographs in Figure 6. For practical applications, it is advantageous that the size of colloidal aggregates such as supraballs be monodisperse. The constituent PS nanospheres in these supraparticles can be more regularly packed into hexagonal configurations when a much larger number of particles are contained in the emulsion droplets or the size of the droplets increases. The shape and size of this nanoreactor can be modified by light and give specific adaptation to the number of local contain for supraparticles.

Since the ordered hexagonal packing of PS particles possesses periodic refractive index variation. We could use the supraball inside the hole template as photonic balls which can reflect the specific wavelength of visible light similar to photonic crystals as it was previously shown [15, 16]. The modification of the diameter or the circular symmetry leads to different aggregation of beads. Another approach of this study could be to coat proteins onto polystyrene nanospheres and then to deposit them into an array of doughnut shape holes as microwells which could provide physical confinement to the spheres [17].

One other good example could be the possibility to create plasmonic ring resonator with an array of silver nanorods. The plasmon resonance can then be modified by changes of the diameter of the nanodoughnut-containing-nanoparticules. Moreover, due to the unique property of an azopolymer surface to immobilize and imprint nanoparticles and biomolecules under the light illumination [18], azopolymer nano-doughnuts can become a general tool for any optical, chemical and biological applications which can be benefited from a functional photoreshape sensing the nanoenvironment. Label-free detection with nanocavities can be implemented with nano-doughnut.

A study of size-dependent properties of similarly shaped nanostructures, such as light localization, diffraction, behavior of trapped particles and photomechanical responses, would be of general interest of nanophotonics [19]. A wide range of sizes, property of flexible photoreshaping, ability to form single and clustered nano-doughnuts, meet all necessary requirements for this study. By choosing the illumination light wavelength in the azopolymer absorption band or far from it, doughnut-shaped nanostructures can switch their behavior and act as flexible reshaping to stable transparent nano-objects, thus performing a wide opportunity of utilization. Furthermore, due to the ability of azopolymer surface relief structures to induce surface plasmon resonances after coating with a metal [20], it is possible to simply produce plasmonic nano-doughnuts that can be useful for various applications including biosensors, solar cells and plasmonic traps. Moreover it was recently demonstrated a controllable preparation of microspheres using geometrically mediated droplet formation in a single mold [21]. Changing the mold geometry can control the size of the microspheres. The size control of doughnut-shape structure can act as a mold to obtain microspheres with a surfactant-free method. Additionally, all presented results can assist theoretical understanding of an incoherent light interaction with azopolymer nanostructures.

3. Conclusion

In summary, we have experimentally demonstrated a possibility of controlled near-field optical modifications of an azopolymer doughnut-shaped nano-object. The photoinduced formation of nanotructures following ny near-field reshaping opens up new opportunities for flexible creation of nano-objects. The process described in this paper has the capability to organize spherical colloids into supported 2D arrays of aggregates that have both positional and orientational orders.

Presented nano-doughnuts can assist any possible application which needs a simultaneous utilization of similar nano-objects with varying sizes. Furthermore, due to the wide topography variation of the azopolymer film, nano-doughnuts can be used both as single nano-objects and as clusters of several nano-doughnuts for performing doughnut-to-doughnut coupling effects. Finally, we emphasize that sufficiently wavelength-dependent nature of an azopolymer response to light pushes forward the utilization of the doughnuts as functional nano-objects and nanoreactors.

Acknowledgements

The authors would like to acknowledge the financial support of CNRS and Région Pays de la Loire. We are grateful to SCIAM laboratory at the University of Angers for the help with obtaining SEM images. Special thanks to A. Diacon for supplying us with polystyrene nanospheres.

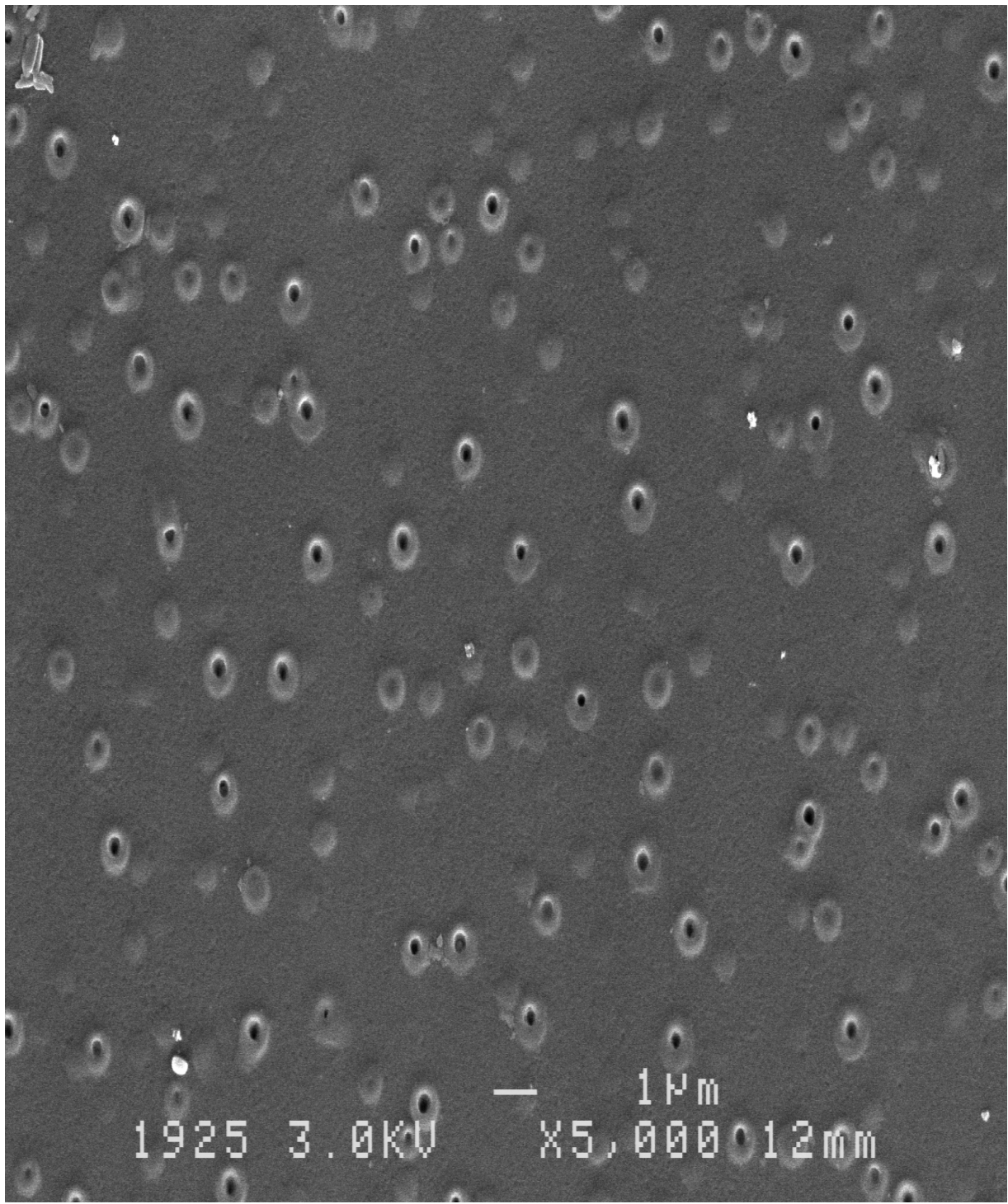

**Figure 1:** Large scale MEB image of the photo-induced nano-well.

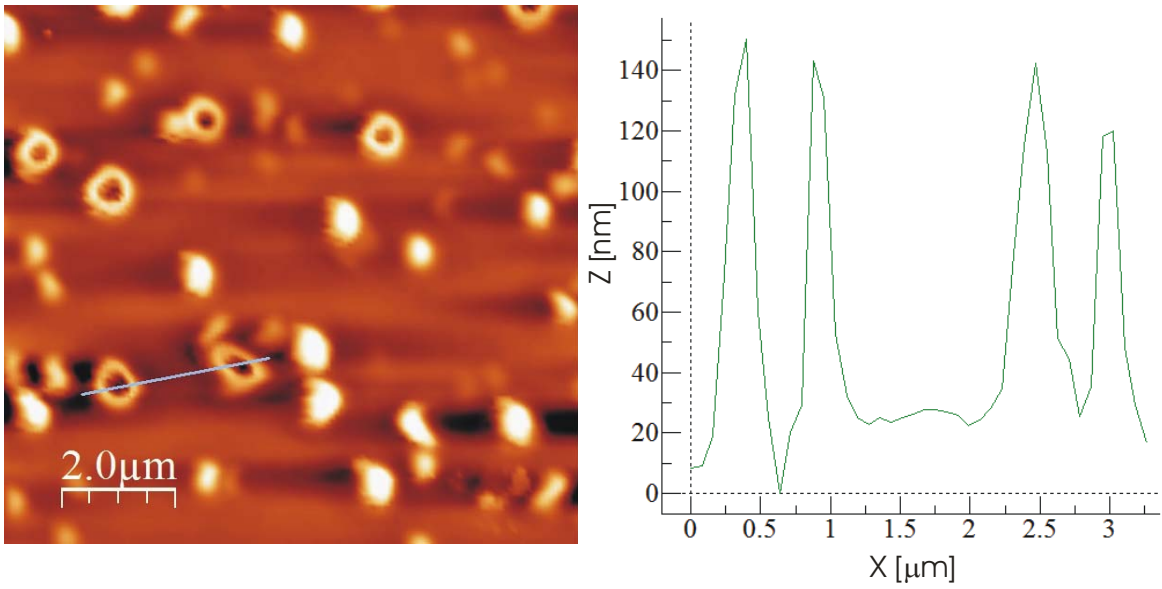

**Figure 2:** Topographic measurement of two doughnut-shaped structures measured with AFM in a contact mode.

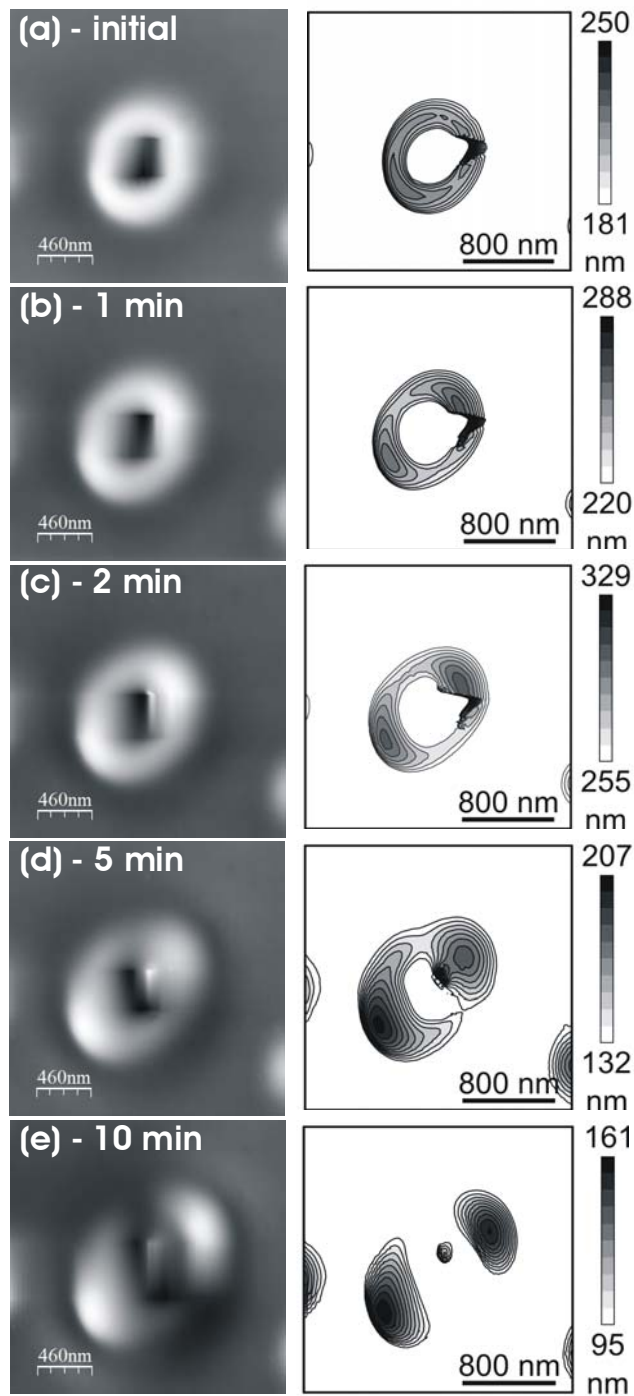

**Figure 3:** Large scale SNOM shear-force image of the photo-induced nano-well after the complete procedure of near-field reshaping. (d-h) Close-up topographic images of the nano-well after certain times of illumination through SNOM probe. The illumination time is specified below each image with contour plot images of the nano-well corresponding to images (d-h). Height color bars are rescaled to better show an upper part of the nano-well structure.

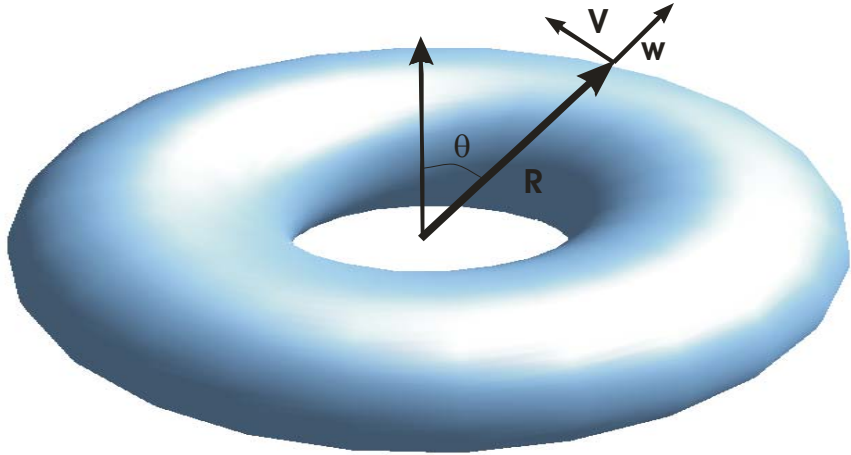

**Figure 4:** Representation of the doughnut-shaped structure with all the parameters used for calculation.

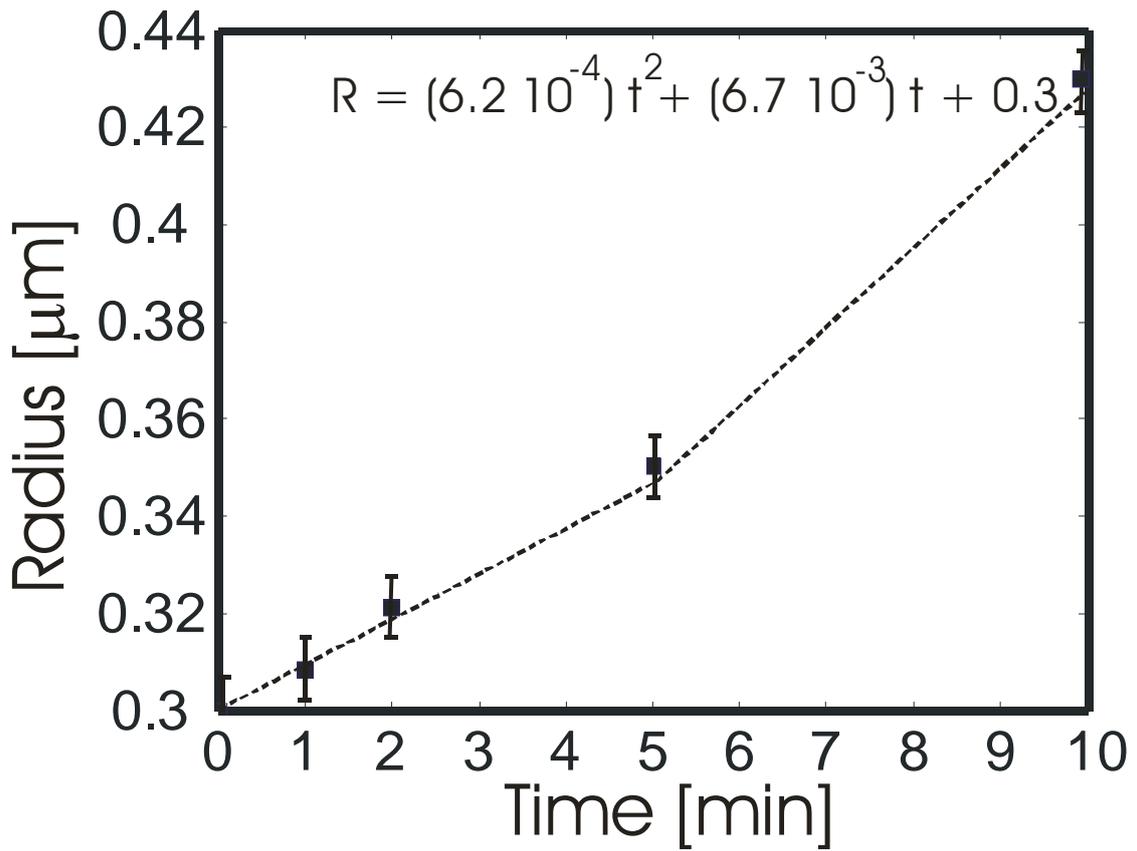

$R = (6.2\ 10^{-4})\ t^2 + (6.7\ 10^{-3})\ t + 0.3$

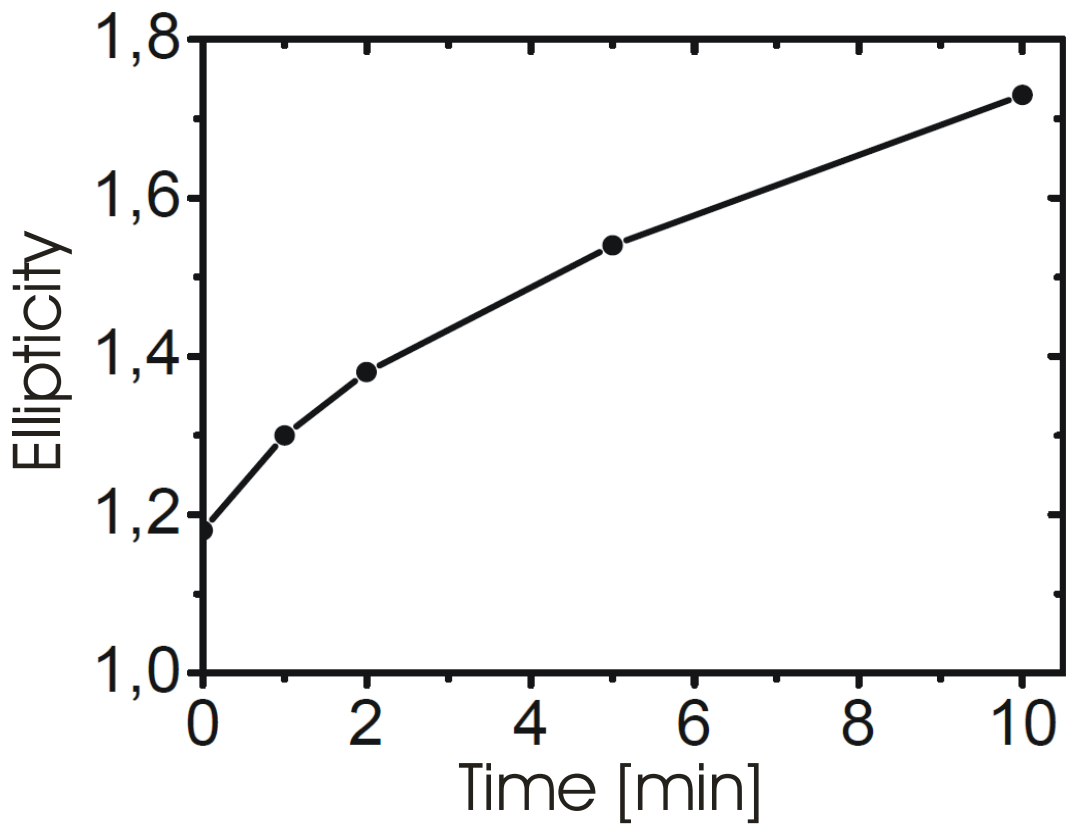

**Figure 5:** Calculated ellipticity factors after the 1st, 3rd, 5th and 10th minutes of the illumination.

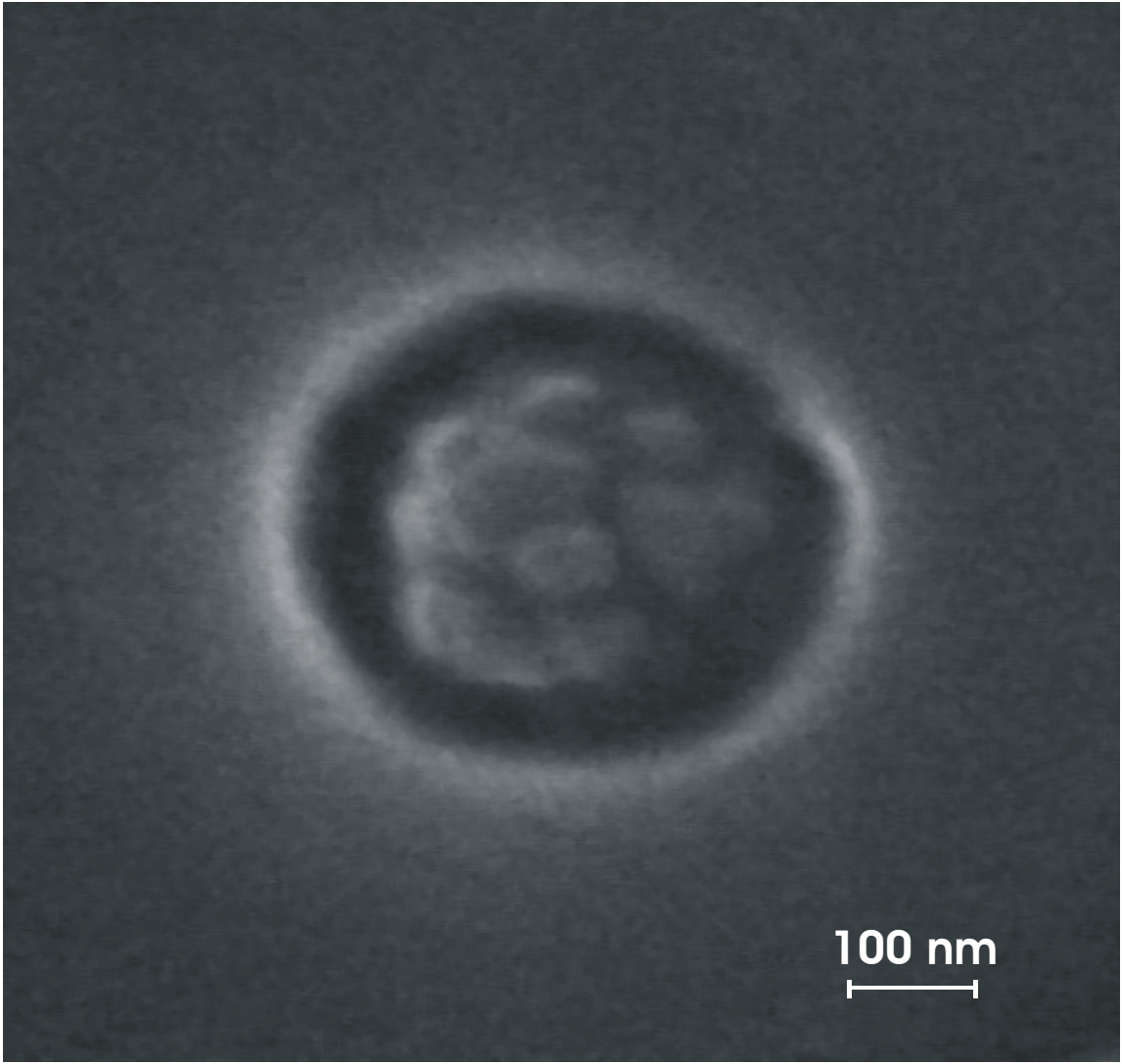

**Figure 6:** SEM image of polysterene nanospheres self-arranging in a typical area of the film with doughnut-shaped nanostructures. The image shows details of polysterene nanospheres trapped as a supraparticle in the azopolymer doughnut-shaped nanostructures. Image was recorded with high control voltage (15 kV) that makes doughnuts to appear as slightly visible rings around trapped nanospheres.